\documentclass[apjl]{emulateapj}

\usepackage[colorlinks=true,citecolor=blue,linkcolor=blue]{hyperref}

\shorttitle{UV Luminosity Functions}
\shortauthors{Oesch et~al.}

\usepackage{color}

\begin{document}

\title{The Evolution of the UV Luminosity Function from $z\sim 0.75$ to $z\sim 2.5$ using HST ERS WFC3/UVIS Observations
\altaffilmark{1}}

\altaffiltext{1}{Based on data obtained with the \textit{Hubble Space Telescope} operated by AURA, Inc. for NASA under contract NAS5-26555. }

\author{P. A. Oesch\altaffilmark{2},
R. J. Bouwens\altaffilmark{3,4}, 
C. M. Carollo\altaffilmark{2}, 
G. D. Illingworth\altaffilmark{3}, 
D. Magee\altaffilmark{3}, \\
M. Trenti\altaffilmark{5}, 
M. Stiavelli\altaffilmark{6},
M. Franx\altaffilmark{4}, 
I. Labb\'{e}\altaffilmark{7},
P. G. van Dokkum\altaffilmark{8}
}

\altaffiltext{2}{Institute for Astronomy, ETH Zurich, 8092 Zurich, Switzerland; poesch@phys.ethz.ch}
\altaffiltext{3}{UCO/Lick Observatory, University of California, Santa Cruz, CA 95064}
\altaffiltext{4}{Leiden Observatory, Leiden University, NL-2300 RA Leiden, Netherlands}
\altaffiltext{5}{University of Colorado, Center for Astrophysics and Space Astronomy, 389-UCB, Boulder, CO 80309, USA}
\altaffiltext{6}{Space Telescope Science Institute, Baltimore, MD 21218, United States}
\altaffiltext{7}{Carnegie Observatories, Pasadena, CA 91101, Hubble Fellow}
\altaffiltext{8}{Department of Astronomy, Yale University, New Haven, CT 06520}

\begin{abstract}

We present UV luminosity functions (LFs) at 1500 \AA\ derived from the HST
Early Release Science WFC3/UVIS data acquired over $\sim50$ arcmin$^2$ of
the GOODS-South field. The LFs are determined over the entire redshift
range $z=0.75-2.5$ using two methods, similar to those used at higher
redshifts for Lyman Break Galaxies (LBGs): (1) 13-band UV+optical+NIR
photometric redshifts to study galaxies in the range $z=0.5-2$ in three
bins of $dz=0.5$, and (2) dropout samples in three redshift windows
centered at $z\sim1.5$, $z\sim 1.9$, and $z\sim 2.5$.  The characteristic
luminosity dims by 1.5 mag from $z=2.5$ to $z=0.75$, consistent with 
earlier work.  However, the other Schechter function parameters, the
faint-end slope and the number density, are found to be remarkably constant
over the range $z=0.75-2.5$. Using these LF determinations we find the UV
luminosity density to increase by $\sim1.4$ dex according to
$(1+z)^{2.58\pm0.15}$ from $z\sim0$ to its peak at $z\sim2.5$.  Strikingly,
the inferred faint-end slopes for our LFs are all steeper than
$\alpha=-1.5$, in agreement with higher-redshift LBG studies. Since the
faint-end slope in the local universe is found to be much flatter with
$\alpha\simeq-1.2$, this poses the question as to when and how the expected
flattening occurs. Despite relatively large uncertainties, our data
suggest $\alpha\simeq-1.7$ at least down to $z\sim1$. These new
results from such a shallow early dataset demonstrate very
clearly the remarkable potential of WFC3/UVIS for the thorough characterization of galaxy evolution
over the full redshift range $z\sim0.5$ to $z\sim3$. 
\end{abstract}

\keywords{galaxies: evolution ---  galaxies: high-redshift --- galaxies: luminosity function, mass function}

\section{Introduction}

The UV Luminosity Function (LF) of galaxies is a key diagnostic for
establishing the contributions of galaxies of different luminosities
and masses to the cosmic star-formation rate (SFR) density. In
addition, in combination with other measures, like the galaxy mass
function, it also provides essential clues as to the causes for the
slow down in star-formation of galaxies in more recent times since $z\sim2$.

The UV LF has been studied very extensively at redshifts $2\lesssim
z\lesssim6$, but at lower redshifts ($z\la2$), similar deep, high
resolution data has not been available. To date the $z<2$ UV LF has been measured
from GALEX data using the dropout technique \citep[e.g.][]{Burgarella06,Ly09,Haberzettl09}, and also using spectroscopic or
photometric redshifts \citep[e.g.][]{Arnouts05,Wyder05,Budavari05}.  The
main limitations of the GALEX results are (1) the very wide point-spread
function (PSF; $\sim5$\arcsec\ FWHM), which may cause blending of the UV
light from several sources and makes the identification of optical
counterparts relatively difficult, and (2) the limited depth ($\sim25$
mag).

With the installation of the Wide-Field Camera 3 (WFC3) on the Hubble
Space Telescope (HST), the efficient coverage of the electromagnetic
spectrum at high spatial resolution ($\lesssim$0\farcs15) has been extended
to the UV and to the IR.  The new IR capability allowed us (and other
groups) to perform the first robust measurements of the UV LF at $z\sim7-8$
in the first-epoch data of the HUDF09
\citep[e.g.][]{Oesch10c,Bouwens10a,McLure10,Bunker09}.

The high sensitivity UV imaging capability provided by the WFC3/UVIS
channel similarly allows us to derive the UV LF at redshifts below
$z\sim2.5$, where the $\sim$1500 \AA\ UV flux from galaxies could not
be probed efficiently with previous high resolution cameras. 
While the depth of the current ERS WFC3/UVIS UV data \citep{Windhorst10} is less than that of
the WFC3/IR data, the present depths still
significantly exceed that from any previous UV data set.

We use two
different methods for our LF determinations: (1) 13-band UV+optical+NIR photometric redshifts to
study galaxies in the range $z\sim0.5-2$, and (2) the UV dropout
technique, applied to the UVIS filter set, which allows for the
selection of star-forming galaxies at $z\sim1.3$ to $z\sim2.8$. The latter
dropout samples are similar to the recent selections of
\citet{Hathi10}.

In Section \ref{sec:data} we describe the data, the photometric
redshift estimates and the dropout selection. This is followed in Section
\ref{sec:LF} by our constraints on the UV LF between $z = 0.5 - 2.8$, and
we end with a short discussion on the evolution of the UV LF parameters
and luminosity density in Section \ref{sec:discussion}.

Throughout this letter, we adopt $\Omega_M=0.3, \Omega_\Lambda=0.7, H_0=70$
kms$^{-1}$Mpc$^{-1}$, i.e. $h=0.7$. Magnitudes are given in the AB system
\citep{Oke83}.

\section{Data and Sample Selection}
\label{sec:data}

\subsection{ERS Data Set}

The Early Release Science (ERS) UV data cover a total of $\sim$50
arcmin$^2$ of the GOODS-South fields \citep{Giavalisco04b} in
$4\times2$ pointings of WFC3/UVIS in three filters ($UV_{225}$ --
F225W, $UV_{275}$ -- F275W, $U_{336}$ -- F336W). Each pointing was
covered with five orbits, equivalent to 2778~s, 5688~s, and 5688~s
integrations in $UV_{225}$, $UV_{275}$, and $U_{336}$, respectively.
The observations in each band were split into three ($U_{336}$) or
four ($UV_{225/275}$) exposures to facilitate cosmic-ray (and bad
pixel) removal \citep[see][for more information on the data]{Windhorst10}. The RMS maps were scaled to the proper
noise level, as measured in circular apertures appropriate for the galaxies under study.  
Using circular apertures of 0\farcs2 radius, the
measured $5\sigma$ point source sensitivities in all three filters are
26.8-26.9 mag.

After masking areas with bad sampling and large residuals from cosmic
ray hits the final area used in this study with both ACS and WFC3/UVIS coverage
is $\sim47$ arcmin$^2$.

The reduced UVIS images were registered to the GOODS ACS
B-band frames (version 2; M. Giavalisco and the GOODS Team, 2010, in
preparation\footnote{http://archive.stsci.edu/pub/hlsp/goods/v2/}) and
drizzled to the same pixel scale of 0\farcs03. The most recent
in-flight geometric distortion solutions (02/2010) were essential for
obtaining good alignment.  Sources were detected in the ACS B-band
image using SExtractor \citep{Bertin96}, resulting in 6683 detections    
down to 28 mag.  Colors were measured in dual image mode in isophotal apertures to
maximize the S/N.

Lastly, our catalog was matched against the public GOODS catalogs and
candidate stars flagged. At $z_{850}<25.5$ this identification is
based upon their size (half-light radii $<0$\farcs09) and the SExtractor
stellarity parameter ($>0.8$). Faintward of 25.5, the latter parameter is no
longer reliable, and therefore flagging sources with half-light radii
$<0$\farcs09 was done if the $V-i$ vs. $i-z$ colors of sources were
consistent with the stellar sequence (within 0.15 mag).   From a visual
inspection of the sample of galaxies entering in the LF estimates this has
proven to be a very efficient and complete procedure to identify stars.

\subsection{Photometric Redshift Selections}

In order to derive accurate photometric redshifts, we supplement the high
resolution ACS and WFC3/UVIS data with ground-based IR and with Spitzer
data, using the GOODS-MUSIC catalog (v2) of \citet{Santini09}.  In
particular, the GOODS-MUSIC photometry used here is VIMOS $U$, ACS
$B_{435}/V_{606}/i_{775}/z_{850}$, ISAAC $J, H, K$, and Spitzer IRAC 3.6
and 4.5. The GOODS-MUSIC catalog is 90\% complete down to $z_{850}\leq26$ mag.
Matching this catalog to our B-selected sources, we find 4165 matches
(using a 0\farcs2 error tolerance). The UVIS isophotal fluxes are scaled to total fluxes using the ratio of B-Band ISO flux to B-Band AUTO flux.

Thus, our basic photometry catalog for photo-$z$ estimates consists of 7
filter HST data, complemented with 4 PSF-matched ground based filters, and
the two shortest wavelength IRAC bands. The IRAC 5.8 and 8.0 bands were
omitted, since they are dominated by dust and PAH emission for lower
redshift galaxies. 
Even if galaxies are generally red in $B-z$, we limit our
analysis to $B_{435}<26$ mag to avoid incompleteness corrections due to the
$z_{850}$-band limit. We did not use the WFC3/IR photometry, since it is not
available for all sources to keep our analysis as uniform as possible (and
our tests show that our selections do not change significantly when these
data are included).

Photometric redshifts were estimated from spectral energy distribution (SED)
fitting with ZEBRA \citep{Feldmann06}. The adopted SED set is based on
\citet{Bruzual03} composite stellar populations. In particular, we computed
15 median templates as a function of rest-frame $u-J$ color for galaxies in
the zCOSMOS survey \citep{Lilly07,Lilly09}. Subsequently, these templates
were corrected for systematic offsets with respect to the COSMOS photometry
\citep{Capak07} using ZEBRA, and the major emission lines (Ly$\alpha$,
H$\alpha$, H$\beta$, OII, and OIII) were added using the \citet{Kennicutt98}
relations between UV luminosity and emission line strengths.  Finally, dust
was added to this set of templates using the \citet{Calzetti00} dust law
with $E(B-V) = 0-0.5$, resulting in a final set of 2052 templates.

We used this template set to derive photometric redshifts for our combined
catalog of UVIS and GOODS-MUSIC photometry.  
We compared our estimates with spectroscopic redshift measurements compiled in the GOODS-MUSIC catalog. From a total of 462 sources with reliable spectroscopy (363 at $z<1.5$), we find an accuracy of $\sigma_z = 0.037\times(1+z)$, with only $6.4\%$ of sources having redshift errors larger than $|\Delta(z)|>0.15\times(1+z)$.
Our photometric redshift
estimates are therefore sufficiently precise to derive LFs in bins of $dz =
0.5$.

\begin{figure*}[htbp]
	\centering
		\includegraphics[scale=0.45]{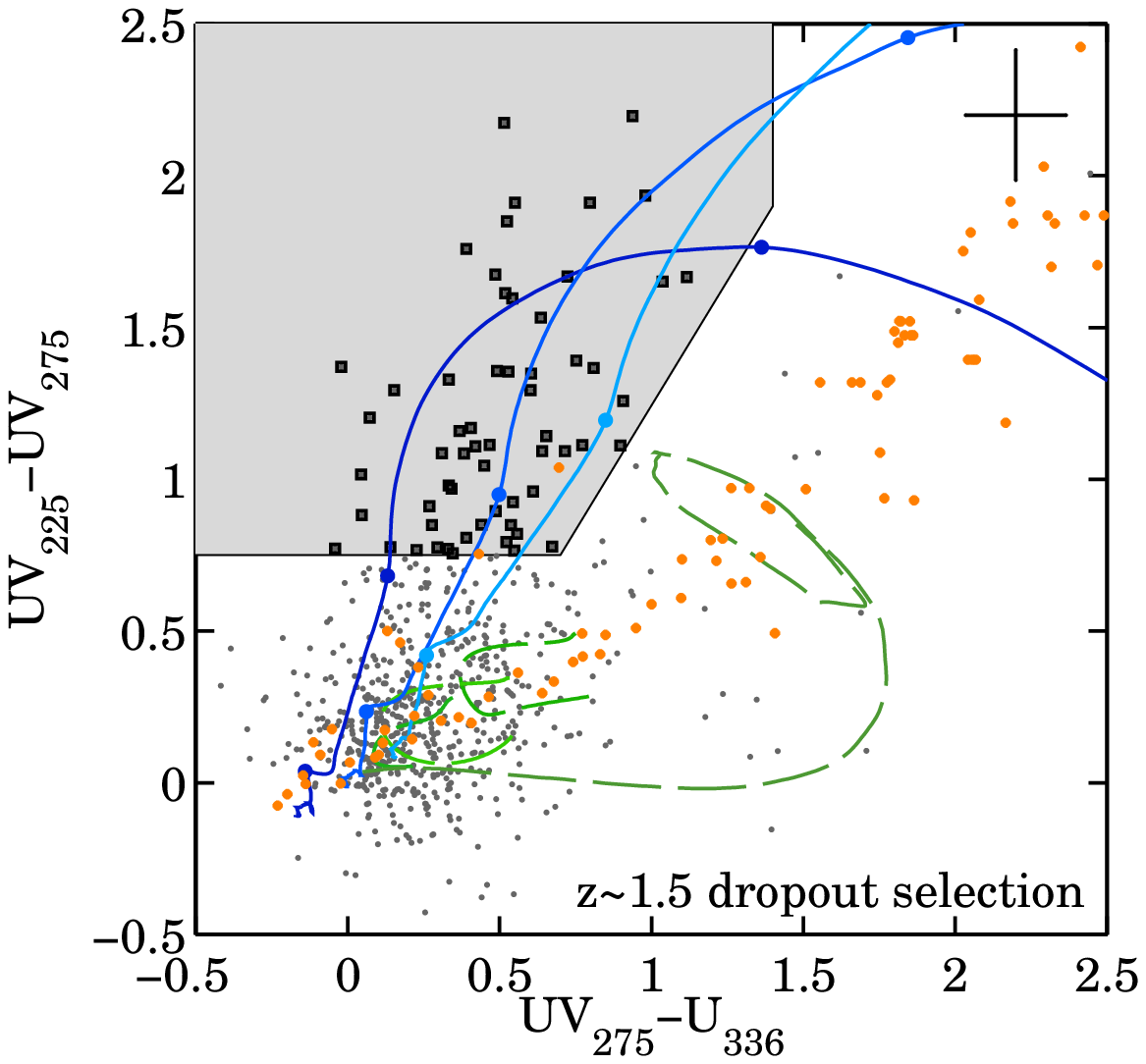}
		\includegraphics[scale=0.45]{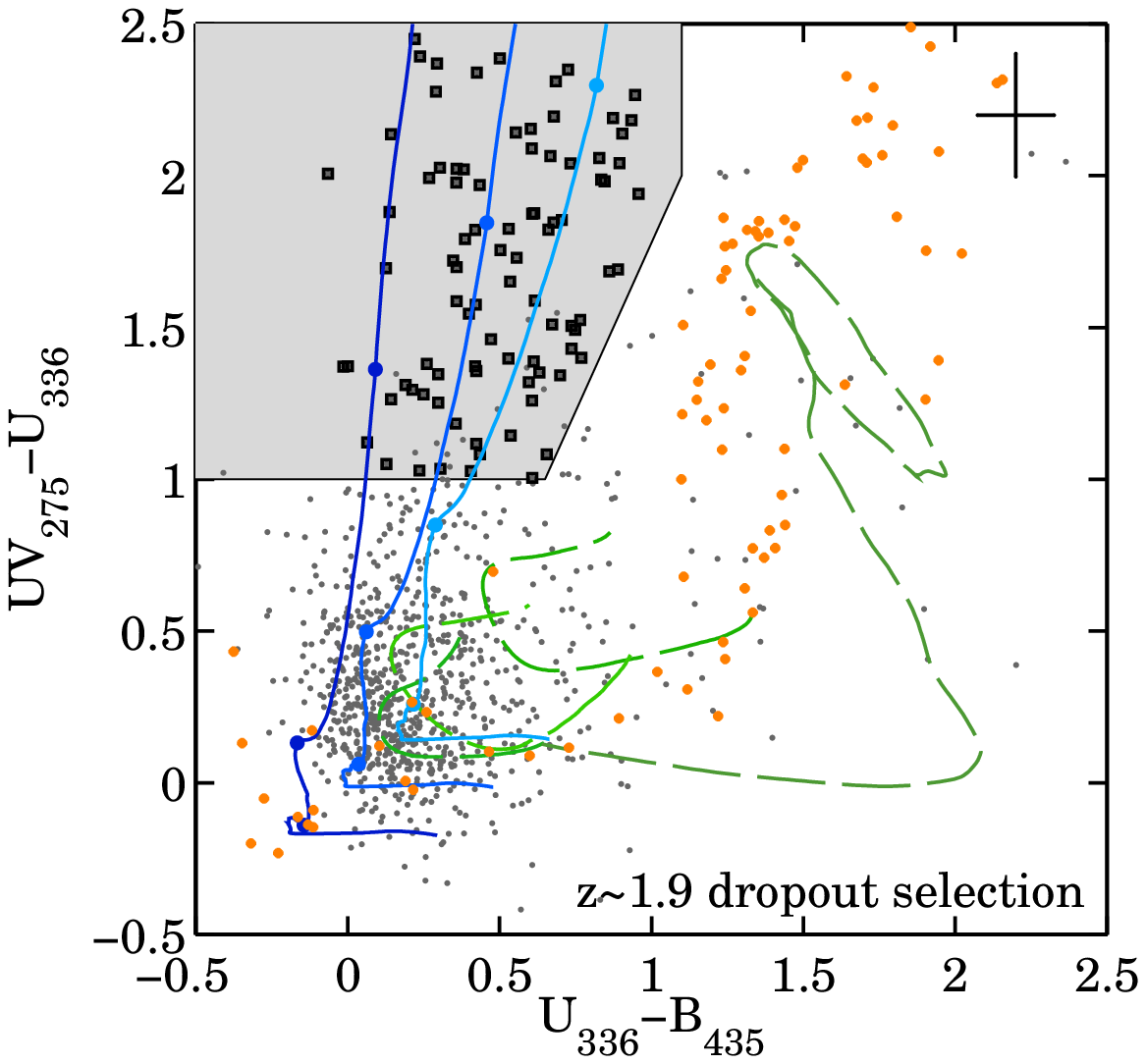}		
		\includegraphics[scale=0.45]{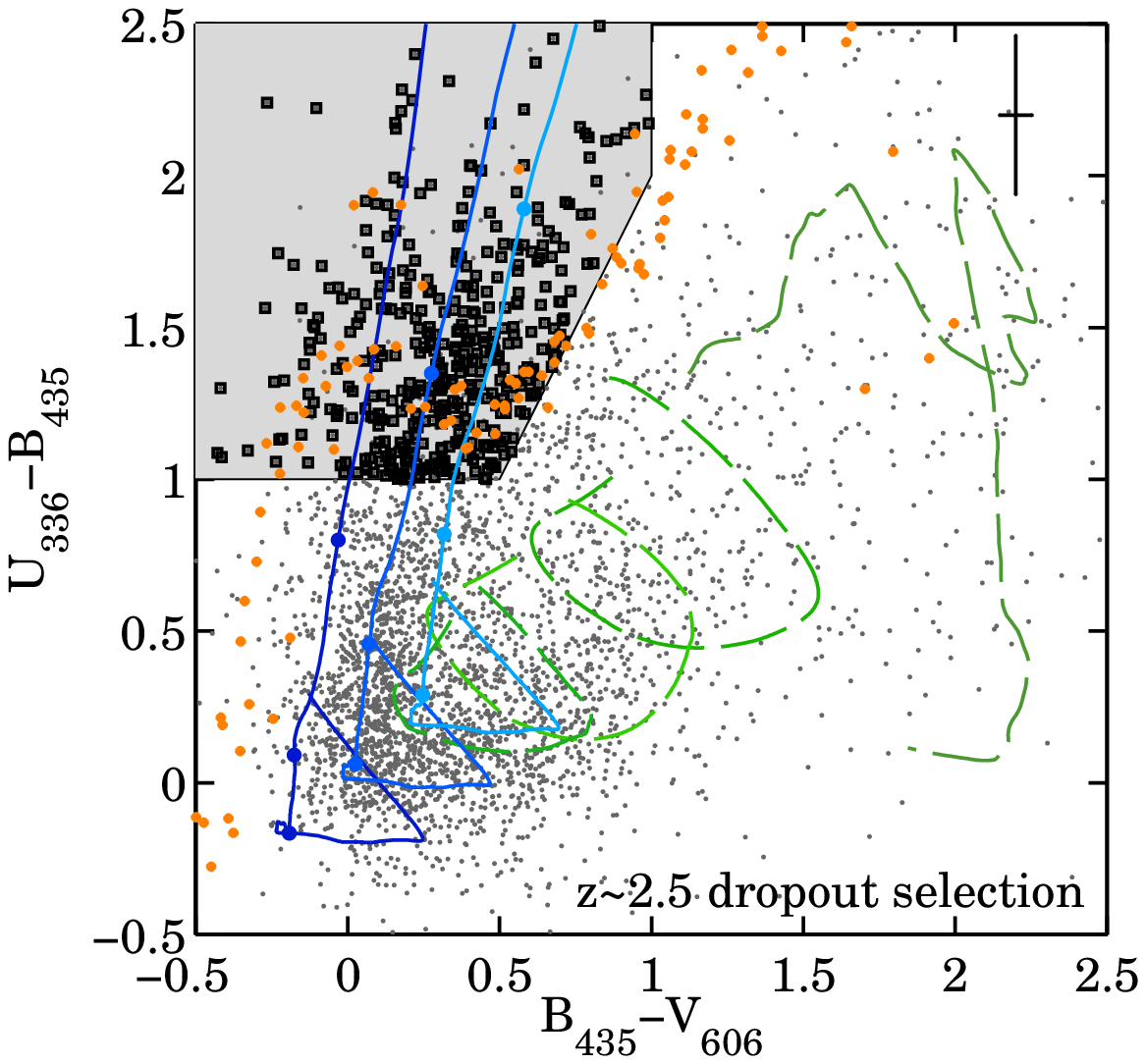}		
	\caption{The color selection of UV-dropouts at redshifts $z\sim1.5,~1.9$,
and $2.5$ (from left to right). Blue lines represent tracks of star-forming
galaxies with dust obscuration of $E(B-V) = 0,~0.15,~0.3$. The green dashed
lines are the expected colors of lower redshift galaxies \citep{Coleman80}.
These tracks extend to $z<1.1$, $z<1.5$, and $z<2$ for the different
figures respectively. The gray shaded regions are the adopted selection
regions for dropout galaxies. The orange points are stars from the
\citet{Pickles98} library.  Stars are removed from our samples
using a combination of the stellarity parameter and  $(V-i)$ vs. $(i-z)$
colors, as described in the text. The light gray dots represent all
galaxies in the catalog down to the respective magnitude limits adopted in
our LF determinations (see section \ref{sec:UVdropLF}), and black squares
indicate our candidates. Fluxes with a detection significance below
$1\sigma$ have been replaced with $1\sigma$ upper limits. Typical errorbars are shown in the upper right.}
	\label{fig:colcol}
\end{figure*}

\subsection{UV Dropout Selection}

Most of the UV LFs at $z>3$ are based on an LBG selection \citep[e.g.][]{Steidel95}. The same
technique can also be applied to the WFC3/UVIS filters, allowing the
selection of star-forming galaxies as UV dropouts.  At $z<3$, the Lyman
break seen in galaxies is primarily the result of neutral hydrogen within
the sources themselves, in significant contrast to the situation at $z>3$,
where the break is also due to Lyman-series absorption from the IGM.

Our selection criteria are based on standard color-color diagrams, which
are shown in Figure \ref{fig:colcol}. In particular, the selection regions
are chosen to identify an unbiased sample of star-forming galaxies with dust
extinction less than $E(B-V)\sim0.3-0.4$ guarding against lower
redshift interlopers.

Specifically, the adopted selection criteria for $UV_{225}$ dropouts are:
\begin{eqnarray*}
& UV_{225}-UV_{275}>0.75~~\wedge &  \\
&UV_{225}-UV_{275} > 1.67(UV_{275}-U_{336})-0.42~~\wedge & \\
&-0.5<(UV_{275}-U_{336})<1.4~~\wedge &\\
&S/N(UV_{275})>5&
\end{eqnarray*}
This selects galaxies between $z\sim1.3-1.7$, and results in 72 candidates down to $U_{336} = 26.5$.

Similarly, for $UV_{275}$-dropouts, the selection criteria are:
\begin{eqnarray*}
& UV_{275}-U_{336}>1 ~~\wedge &  \\
& UV_{275}-U_{336}>2.2(U_{336}-B_{435})-0.42 ~~\wedge & \\
& -0.5<(U_{336}-B_{435})<1.1 ~~\wedge &\\
&S/N(U_{336})>5 ~~\wedge~~ S/N(UV_{225})<2&
\end{eqnarray*}
selecting galaxies between $z\sim1.7-2.1$, with 131 candidates down to $B_{435} = 26.5$.

And finally,  $U_{336}$-dropouts are selected with:
\begin{eqnarray*}
& U_{336}-B_{435} > 1 ~~\wedge &  \\
& U_{336}-B_{435} > 2(B_{435}-V_{606}) ~~\wedge & \\
& -0.5<(B_{435}-V_{606}) < 1 ~~\wedge &\\
&S/N(B_{435})>5 ~~\wedge~~ S/N(UV_{225/275})<2&
\end{eqnarray*}
returning 434 sources down to $V_{606}=26.5$ with expected redshifts
in the range $z\sim2.2-2.8$.  When applying these criteria, all fluxes
below their $1\sigma$ limits (in the dropout band) are replaced with the
respective $1\sigma$ upper limits.

For the $U_{336}$-dropouts, the stellar sequence runs exactly through the
selection window (see Figure \ref{fig:colcol}). However, stars
are removed prior to candidate selection using the same identification
scheme as for the photo-$z$ sample, resulting in a total of 40 excluded
stars. From a final visual inspection of all sources no additional stars
were found. A small number of clear diffraction spikes and spurious
detections on the wings of large low-redshift galaxies were removed by
hand.

Using our photometric redshifts, we estimate the low-redshift interloper
fraction of these UV-dropout samples to be very small. In particular, 
only 7\% of the sources in the $U_{336}$ dropout sample have
$z_{phot}<1.75$. 
Similarly, only 4\% and 8\% of the sources in our $UV_{275}$- and $UV_{225}$-dropout samples are interlopers, i.e., have 
$z_{phot}<1.25$ and $z_{phot}<1$, respectively.
Only sources down to the limiting magnitudes of our UV LFs
are considered in this estimate (see section
\ref{sec:UVdropLF}).

\section{Results}
\label{sec:LF}

\subsection{Photometric Redshift LF Results}

The UV luminosity functions at 1500~\AA\ in the range $z\sim0.5-2$ are
measured in three bins of photometric redshift of width $dz=0.5$. The
absolute magnitudes are computed from the $UV_{275}$, $U_{336}$ and
$B_{435}$ photometry for sources in the redshift range $z=0.5-1$,
$z=1-1.5$, and $z=1.5-2$, respectively. K-corrections are derived from the
appropriate best-fit templates.

The UV luminosity functions of galaxies are estimated using the standard $V_\mathrm{max}$ estimator \citep{Schmidt68} such that
\[
\phi(M)dM = \sum_i \frac{1}{ C(m_i) V_{\mathrm{max},i} }
\]
where the sum runs over all galaxies in the given redshift and absolute
magnitude bin. We only include galaxies with $>5\sigma$ detections in the
corresponding filter, which results in the need for a completeness
correction factor $C$. Following \citet{Oesch07,Oesch10c}, this is  
determined from Monte-Carlo simulations in which we add artificial galaxies
to the images and detect them with the same SExtractor parameters as used
for the original B-band catalog. The completeness is $>97\%$ to $\leq25$ mag in the UVIS filters, 
after which it drops to 50\% at 25.8 mag, which is our faint-end limit.

\begin{figure*}[tbp]
	\centering
		\includegraphics[scale=0.5]{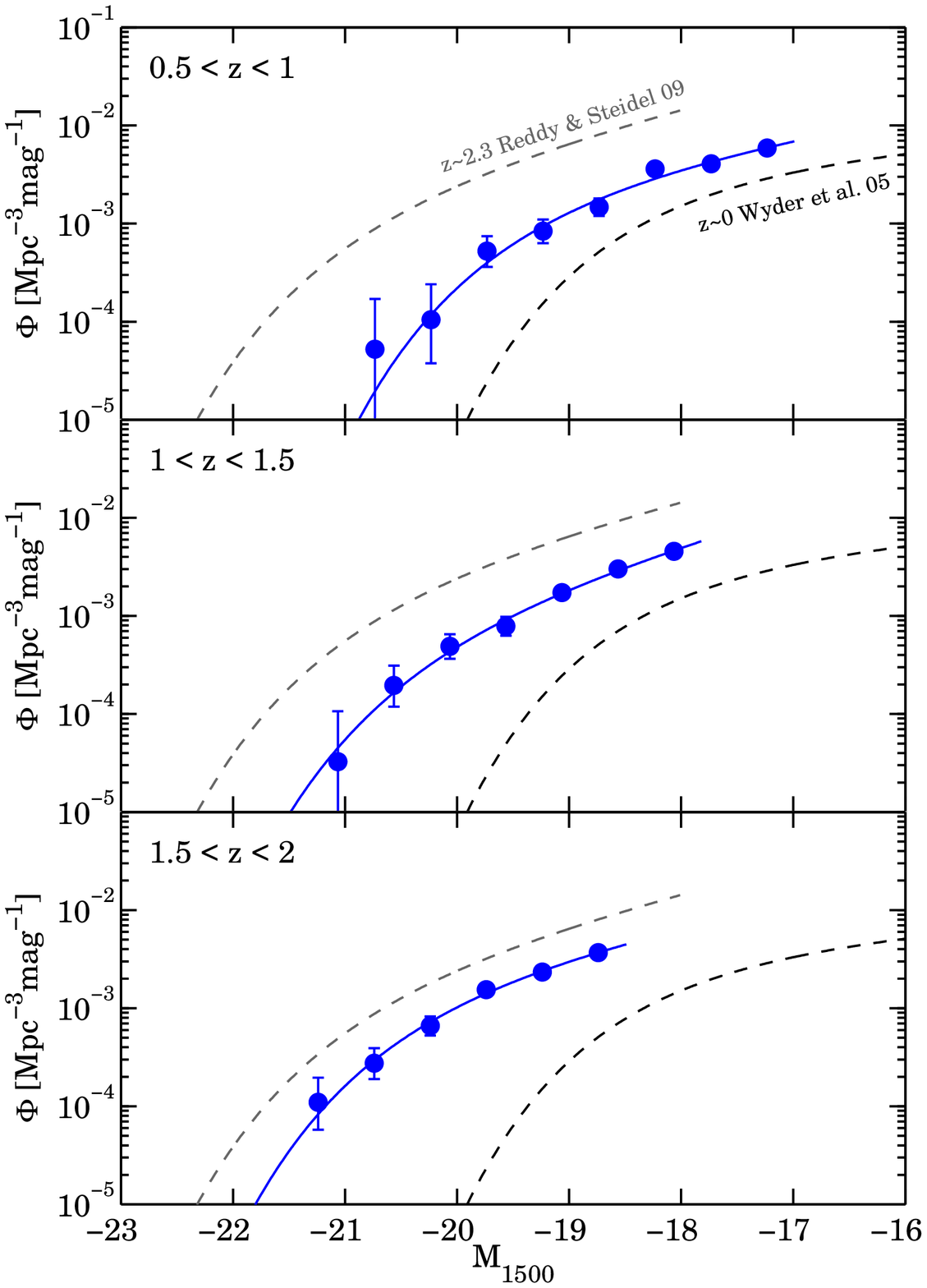}
		\includegraphics[scale=0.5]{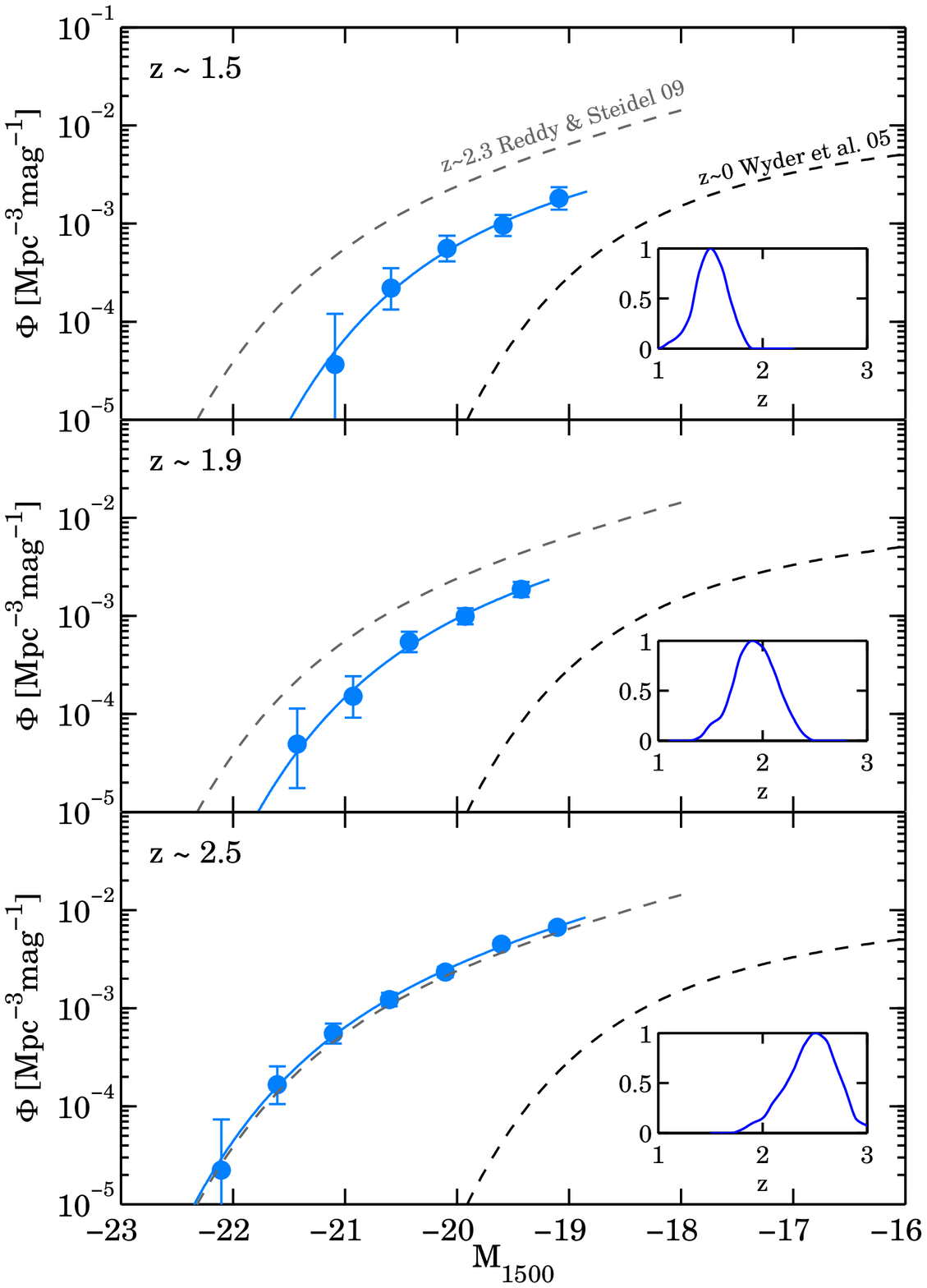}
	\caption{\textit{Left---} The 1500 \AA\ LF in the three different bins of
photometric redshift. The blue data points with errorbars are the step-wise
LFs determined with the $V_\mathrm{max}$ method, and the solid blue lines
represent our best-fit Schechter LFs to the step-wise LFs. The
best-fit parameters are presented in Table \ref{tab:LFresults}. The
black dashed line is the UV LF in the local universe from \citet{Wyder05},
while the gray dashed line is the UV LF between $z\sim1.9-2.7$ from
\citet{Reddy09}.	\textit{Right---} The 1500 \AA\ LF derived from the LBG
samples together with their best-fit Schechter LFs. From top to
bottom: $UV_{225}$-, $UV_{275}$-, and $U_{336}$-dropout results. Again, the local
$z\sim0$ 
and the $z\sim2.3$ LFs are shown for reference as black and gray dashed lines
respectively. The insets show the expected redshift distributions of the
dropout galaxies. This is estimated from the effective volumes and the
best-fit Schechter LFs.}
	\label{fig:UVLF}
\end{figure*}

Figure \ref{fig:UVLF} shows the resulting luminosity functions. The best-fit
Schechter parameters \citep{Schechter76} are tabulated in Table \ref{tab:LFresults}. 
While we sample the faint end rather well, with data that reaches $\sim2$
mag fainter than $M_*$, the relatively small area covered by the ERS data
results in some uncertainty on the cutoff at bright magnitudes.
Nevertheless, our results are in good agreement with the LFs determined
from GALEX by \citet{Arnouts05} and effectively validate those results in
data with $\sim50$ times higher spatial resolution.

\begin{deluxetable*}{ccccc}
\tablecaption{UV Luminosity Function Parameters\tablenotemark{$\dagger$}\label{tab:LFresults}}
\tablewidth{0 pt}
\tablecolumns{5}
\tablehead{\colhead{$z$} & No. of Sources & \colhead{$\log_{10}\phi_*$ [Mpc$^{-3}$mag$^{-1}$]} &\colhead{$M_*$}  &\colhead{$\alpha$}  }

\startdata

\cutinhead{Photometric redshift samples}

$0.5-1.0$      & 284 &       $-2.52\pm0.31$      & $-19.17\pm0.51$      & $-1.52\pm0.25$ \\
$1.0-1.5$      & 284 &       $-2.90\pm0.25$      & $-20.08\pm0.36$      & $-1.84\pm0.15$ \\
$1.5-2.0$      & 288 &       $-2.63\pm0.23$      & $-20.17\pm0.34$      & $-1.60\pm0.21$ \\

\cutinhead{UV-dropout samples}

$1.5$ ($\Delta z=0.4$)      & 60 &       $-2.64\pm0.33$     & $-19.82\pm0.51$     & $-1.46\pm0.54$ \\
$1.9$ ($\Delta z=0.4$)     & 99 &       $-2.66\pm0.36$     & $-20.16\pm0.52$     & $-1.60\pm0.51$ \\
$2.5$  ($\Delta z=0.6$)    & 403 &       $-2.49\pm0.12$      & $-20.69\pm0.17$     & $-1.73\pm0.11$ 
\enddata

\tablenotetext{$\dagger$}{Derived from $\chi^2$ fits to the step-wise LFs.}

\end{deluxetable*}

\vspace{1cm}

\subsection{Dropout LF Results}
\label{sec:UVdropLF}

The UV LF of the UVIS dropout sample is computed using standard techniques adopted in LBG studies. In particular, the step-wise LFs are derived using 
\[
\phi(M[m_i,\bar{z}])dM = N_i/V_\mathrm{eff}(m_i)
\]
where $N_i$ are the number of galaxies observed in the magnitude bin $i$, and the effective volume is given by:
\[
V_\mathrm{eff}(m_i) = \int_0^\infty  S(m_i,z)C(m_i) \frac{dV}{dz} dz
\]
The selection efficiency $S$ and completeness $C$ are again estimated from
simulations of artificial galaxies inserted into the observed images.  The
inserted galaxies are chosen to have a log-normal size distribution,
using an extrapolation of the higher redshift LBG size scaling of $(1+z)^{-1}$
\citep{Oesch10b,Bouwens04a,Ferguson04}. The colors are set according to the
$z\sim2.5$ UV continuum slopes derived in \citet{Bouwens09b} \citep[see
also][]{Reddy08}, with additional IGM absorption according to
\citet{Madau95}.

Due to the lower sensitivity of the UV data with respect to the optical,
red colors across the continuum breaks cannot be accurately measured for
galaxies fainter than $B\sim26$. Therefore only sources with $V<26.3$,
$B<25.5$, and $U_{336}<25.4$ are used to compute the LFs, ensuring that the
selection efficiency is still larger than $50\%$ for $U_{336}$--,
$UV_{275}$--, and $UV_{225}$--dropouts, respectively. This results in 403, 99,
and 60 sources in the three different dropout samples.

The UV LFs for the UVIS dropouts are shown in the right panels of
Figure \ref{fig:UVLF}. Again, a significant increase at the bright end
of the UV LF is seen from lower to higher redshift. Due to the modest
size of the present $UV_{225}$-- and $UV_{275}$--dropout samples, the
Schechter parameters we derive at $z\sim1.5$ and $z\sim1.9$ are still
somewhat uncertain (see Table \ref{tab:LFresults}). The $z\sim2.5$
$U_{336}$--dropout LF, by contrast, is much better constrained from
our data, extending $\sim2$ mag fainter than $M_*$ and with a
faint-end slope $\alpha$ established to within $\pm0.11$. Our
$z\sim2.5$ LF is in excellent agreement with the $z\sim2.3$ UV LF of
\citet{Reddy09}.

Note that due to the limited area of the present survey, our results are 
dominated by cosmic variance, which we estimate to be about $16-20$\%
\citep{Trenti08}.

\begin{figure}[tbp]
	\centering
		\includegraphics[scale=0.55]{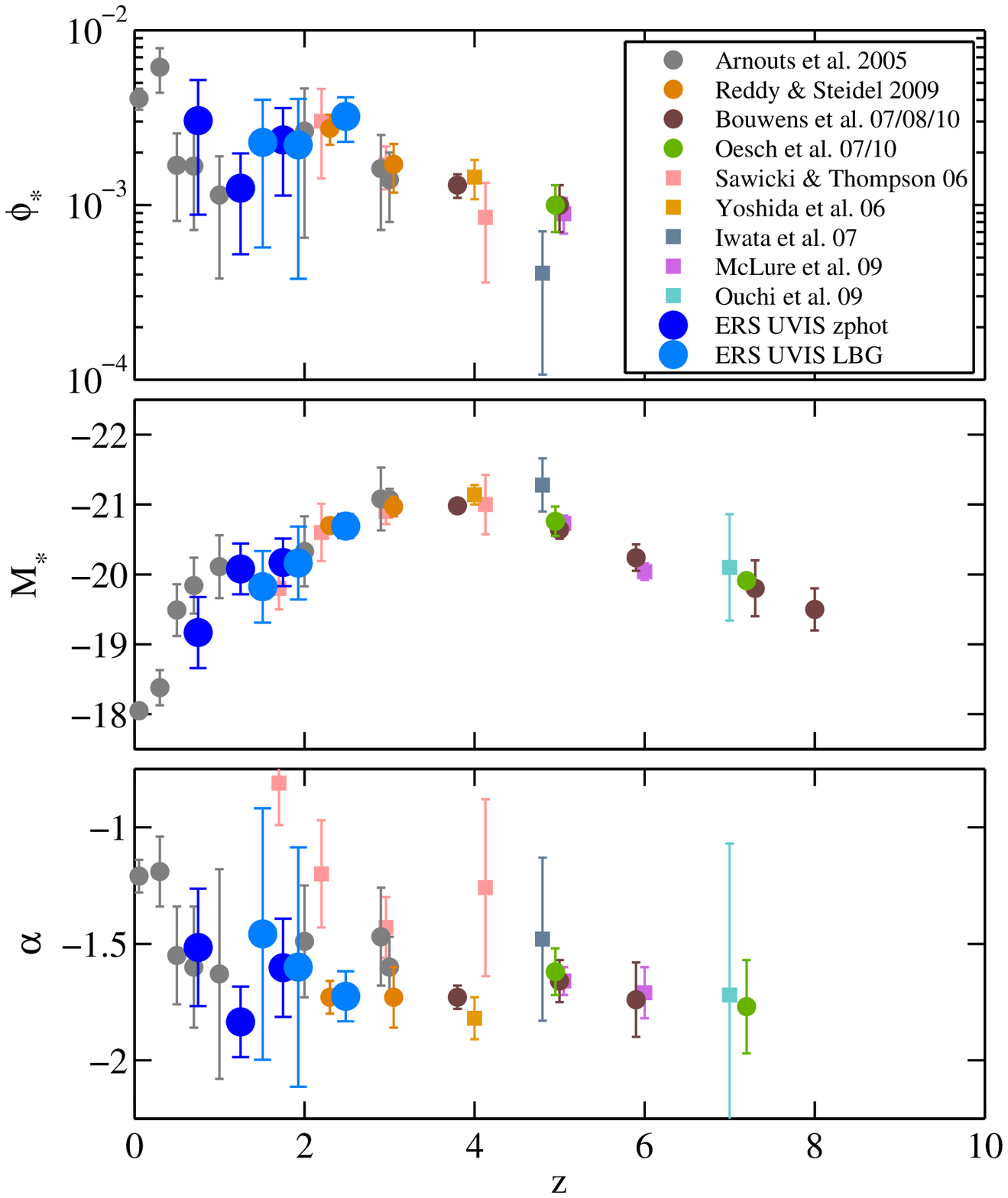}
		\includegraphics[scale=0.55]{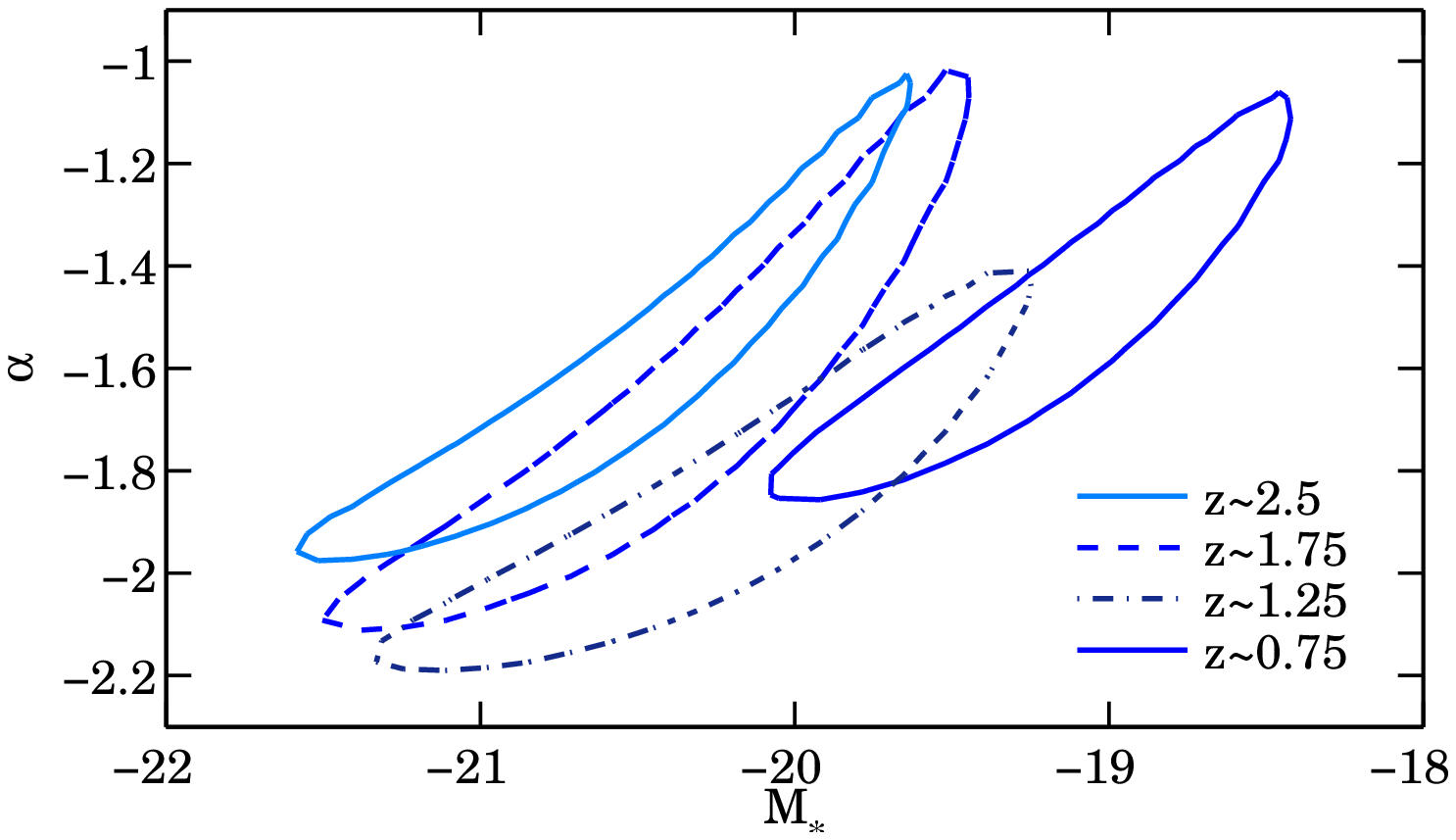}
	\caption{\textit{Top Three Panels --- }The redshift evolution of UV ($\lambda=1500-1700$ \AA) LF
parameters $\phi_*$, $M_*$, and $\alpha$ over the redshift range
$z\sim0-8$. The present determinations are shown with large filled circles.
Of particular note, the uncertainties on the Schechter function parameters
are frequently much smaller at $z>3$ than at lower redshifts. This is a
direct consequence of having very deep HST optical surveys.
WFC3/UVIS should enable similarly-deep future observations in the UV at
$z<3$. Also shown are some notable determinations from
the literature
\citep{Arnouts05,Reddy09,Bouwens07,Bouwens08,Bouwens10a,Oesch07,Oesch10c,Sawicki06,Yoshida06,Iwata07,McLure09,Ouchi09}.
Despite some scatter, most faint-end slope determinations at $z>0.5$ are
very steep, i.e. $\alpha\sim-1.7$.
\textit{Bottom --- } The 68\% error contours on $M_*$ and $\alpha$ over the redshift range $z\sim0.75-2.5$. The contour at $z\sim2.5$ is from our $U_{336}$-dropout sample, the remaining ones are derived from the photometric redshifts. The lower redshift dropout samples result in considerably larger errors and are omitted for clarity.
	}
	\label{fig:LFparams}
\end{figure}

\begin{figure}[tbp]
	\centering
		\includegraphics[scale=0.6]{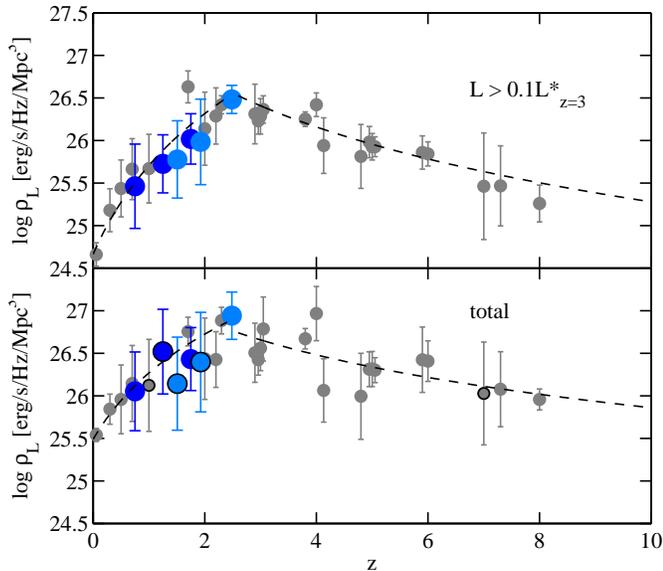}
	\caption{The redshift evolution of the UV ($\lambda\sim1500$ \AA)
luminosity density. In the upper panel, the luminosity functions are
integrated down to a fixed luminosity of $0.1 L^*_{z=3}$ \citep[$L^*$
  as derived at $z=3$ by] []{Steidel99}, while the lower panel shows
the total luminosity density. The
dark and light blue points are derived from our photo-$z$ and the dropout
samples respectively. The gray points are based on the literature
values shown in Figure \ref{fig:LFparams}. The dashed curves are fits
to the data at $z<2.5$ and $z>2.5$ of the form
$\rho_L\propto(1+z)^{\beta_L}$. Errorbars are computed from
Monte-Carlo simulations, not accounting for the covariance between the
Schechter parameters. They are thus expected to overestimate the real
errors. Note, however, that $\alpha$ was constrained to $\alpha>-2$ in order to keep $\rho_L^\mathrm{tot}$ finite. This only significantly affected 5 data points, which are marked with black circles.}
	\label{fig:LD}
\end{figure}

\section{Discussion}
\label{sec:discussion}

The evolution of the Schechter parameters of recent UV LFs in the redshift
range $z\sim0-8$ is shown in Figure \ref{fig:LFparams}.  Our results cover
the whole 4.3 Gyr period from $z\sim0.75$ to $z\sim2.5$.  Interestingly, in
the region of overlap, the LFs we derive from our photometric redshifts and UV dropout selections
are in very good agreement. This suggests that our UV
dropout samples are well-defined and reasonably complete.

In Figure \ref{fig:LFparams}, our results are compared to several
Schechter function estimates from the literature.  Within the
uncertainties, $\phi_*$ is found to be fairly constant over the entire
redshift range, although there is a weak trend towards higher $\phi_*$
at lower redshifts.

The most noteworthy trend is the monotonic fading of the characteristic
luminosity by a factor of $\sim16$ from $z\sim3$ ($M_*$ fades from $-21$ at
$z\sim3$ to $-18$ at $z\sim0$).  
At $z>4$, $M_*$ turns over
and decreases monotonically again towards higher redshift, such that
$M_*(z=8)$ is essentially equal to $M_*(z\sim1)$.

It is very interesting that all our faint-end slope measurements are
steeper than $\alpha \sim -1.5$, even at $z\sim 0.5-1$, 
in agreement with the earlier measurements of \citet{Arnouts05}.
Despite relatively large errorbars, the faint-end slope $\alpha$ is
clearly seen to transition from $\alpha<-1.5$ at $z\gtrsim0.5-2$ to a
flatter $\alpha=-1.2$ in the local universe \citep[but see][who find a steeper slope locally]{Treyer98}.  With the
present data we cannot determine whether this transition occurs
relatively abruptly at low redshift, or smoothly from $z\sim2$ to
$z\sim0$. 


Finally, in Figure \ref{fig:LD} we show the cosmic 1500 \AA\ luminosity
density $\rho_L$ (uncorrected for dust extinction) estimated from the
Schechter function parameters. The luminosity density is shown integrated
to $L>0.1L^*_{z=3}$ and also integrated to zero luminosity.
Clearly, the evolution of $\rho_L$ is mainly driven by $M_*$.
From the peak at $z\sim2.5$ the total UV luminosity drops by 1.5 dex to
$z\sim0$. This decrease is well described by
$\rho_L(z)\propto(1+z)^{\beta_L}$, for which we find $\beta_L(z<2.5) =
2.58\pm0.15$, in good agreement with earlier measurements
\citep[e.g.][]{Schiminovich05,Tresse07}. The evolution at higher redshift
follows $\beta_L(z>2.5) = -1.82\pm0.37$. $z\sim2.5$ thus marks the high
point in the production of the cosmic UV luminosity and is clearly an important epoch for
exploring the properties of galaxies \citep[e.g. see][for
a new technique for selecting galaxies in
this range with simple color criteria]{Cameron10}.

The remarkable potential of the HST WFC3/UVIS camera is clear from these
new results.  This ERS dataset with its initial rather shallow observations
has demonstrated already that WFC3/UVIS is a powerful tool for exploring
galaxy evolution over the redshift range $z\sim0.5$ to $z\sim3$. 
Deeper data will provide key insights into the decline of the cosmic SFRD since $z\sim2$, particularly at the lowest luminosities. For example, a steep LF with $\alpha\sim-1.7$ suggests that star-formation efficiencies and feedback processes are only weakly dependent on halo mass \citep[e.g.][]{Trenti10}.

\acknowledgments{
We thank the GOODS and the GOODS-MUSIC teams for providing their data to the community.
PO acknowledges support from the Swiss National Foundation (SNF). 
This work has been supported by NASA grant NAG5-7697 and NASA grant HST-GO-11563.
}

\vskip0.2cm
Facilities: \facility{HST(ACS/WFC3), VLT(ISAAC), VLT(VIMOS), Spitzer(IRAC)}.

\bibliographystyle{apj}

\end{document}